\documentclass[%
 aip,
 apl,
 amsmath,amssymb,
 reprint,%
]{revtex4-1}

\usepackage{soul}

\usepackage{textcomp}

\usepackage{graphicx}
\usepackage{dcolumn}
\usepackage{bm}

\usepackage{textcomp}
\usepackage{epstopdf}
\usepackage{braket}

\usepackage{color}

\usepackage{hyperref}

\begin{document}

\preprint{AIP/123-QED}

\title[]{Gain dynamics in a heterogeneous terahertz quantum cascade laser}

\author{C.G. Derntl}
 \email{christian.g.derntl@tuwien.ac.at}
\affiliation{%
Photonics Institute, TU Wien, Campus Gusshaus, 1040 Vienna, Austria.
}%
\author{G. Scalari}%
\affiliation{%
Institute for Quantum Electronics, ETH Z{\"u}rich, Auguste-Piccard-Hof 1, 8093 Z{\"u}rich, Switzerland.
}%

\author{D. Bachmann}%
\thanks{Now with Crystalline Mirror Solutions GmbH, Austria.}
\affiliation{%
Photonics Institute, TU Wien, Campus Gusshaus, 1040 Vienna, Austria.
}%

\author{M. Beck}%
\affiliation{%
Institute for Quantum Electronics, ETH Z{\"u}rich, Auguste-Piccard-Hof 1, 8093 Z{\"u}rich, Switzerland.
}%
\author{J. Faist}%
\affiliation{%
Institute for Quantum Electronics, ETH Z{\"u}rich, Auguste-Piccard-Hof 1, 8093 Z{\"u}rich, Switzerland.
}%

\author{K. Unterrainer}
\affiliation{%
Photonics Institute, TU Wien, Campus Gusshaus, 1040 Vienna, Austria.
}%
\affiliation{%
Center for Micro- and Nanostructures, TU Wien, Campus Gusshaus, 1040 Vienna, Austria.
}%
\author{J. Darmo}
 \email{juraj.darmo@tuwien.ac.at}
\affiliation{%
Photonics Institute, TU Wien, Campus Gusshaus, 1040 Vienna, Austria.
}%

\date{\today}

\begin{abstract}
The gain recovery time of a heterogeneous active region terahertz quantum cascade laser is studied by terahertz-pump – terahertz-probe spectroscopy. The investigated active region, which is based on a bound-to-continuum optical transition with an optical phonon assisted extraction, exhibits a gain recovery time in the range of 34$\,$-$\,$50$\,$ps dependent on the operation condition of the laser. The recovery time gets shorter for stronger pumping of the laser while the recovery dynamics slows down with increasing operation temperature. These results indicate the important role of the intracavity light intensity for the fast gain recovery. 
%
\end{abstract}

															%
\maketitle

Semiconductor terahertz lasers based on the quantum cascade design, so called THz QCLs, have become an established THz technology.\cite{doi.org/10.1088/1361-6463/50/4/043001} Besides the generation of spectrally bright THz light, todays’ THz QCL can be used as frequency comb source,\cite{Markus2018} as transceiver at THz frequencies for sensing and communication,\cite{Dean:11} or as a source of tailored coherent pulses generated on demand.\cite{Jukam2009,Bachmann:16} Although the physics of QCLs is well understood\cite{FaistBookQCL2013} and can be modeled to a satisfactory level,\cite{doi:10.1063/1.4863665} their performance is still not under full control.
In experimental studies, discrepancies between the theoretical and actual performance of THz QCLs were observed that were attributed to deviations of the fabricated devices from the design model. To address these issues and to identify their origin THz time-domain spectroscopy (THz-TDS) has been applied to access the internal processes in QCLs,\cite{Kroell2007, doi:10.1063/1.2970046} which allowed to study the spectral gain curve at all operation points of a QCL (hence even above the lasing threshold),\cite{Kroell2007, doi:10.1063/1.3158592,doi:10.1063/1.4901316, doi:10.1063/1.3553021} the gain clamping dynamics,\cite{Jukam2009} and the gain induced dispersion.\cite{Parz:09, doi:10.1063/1.4969065, doi.org/10.1364/OPTICA.3.001362} This flexible spectroscopic tool helped to identify individual gain degradation mechanisms,\cite{doi:10.1063/1.2729992} and provided direct access to the gain recovery dynamics.\cite{doi:10.1063/1.4942452, Freeman2013}
Fast gain dynamics, which is described by a short characteristic time -- the gain recovery time (GRT) -- is important for the high speed modulation of THz QCLs and for the formation and the sustainability of THz pulses in QCLs. Thereby the gain modulation and the pulse dispersion are key issues for the active/passive mode-locking \cite{doi.org/10.1038/nphoton.2011.49, doi:10.1063/1.4765660} of lasers. All present studies of the gain dynamics have been performed on QCLs with the bound-to-continuum (BTC) design that features a relatively narrow gain spectrum of $\sim\,$200$\,$GHz and have reported a GRT in the range of 15$\,$-$\,$26$\,$ps without the phonon extraction scheme.\cite{doi:10.1063/1.4942452, Freeman2013, Green2009,  Markmann:17} Following the recent successful demonstration of a THz QCL with octave spanning gain,\cite{Roesch2014} heterogeneous QCL active regions composed from several different quantum cascades structures become increasingly important.\cite{Turcinkova2011} While for standard active regions of THz QCLs, a single cascade design (typically 3$\,$-$\,$8 quantum wells) is repeated hundreds of times to form a quantum cascade heterostructure (QCH), in a heterogeneous active region several heterostructures with different design details are stacked to form a broadband QCL. As a consequence, the different heterostructures can feature different gain dynamics due to tiny variations in the sequence of quantum wells and barriers.

\begin{figure}[h!]
\centering\includegraphics[width=\linewidth]{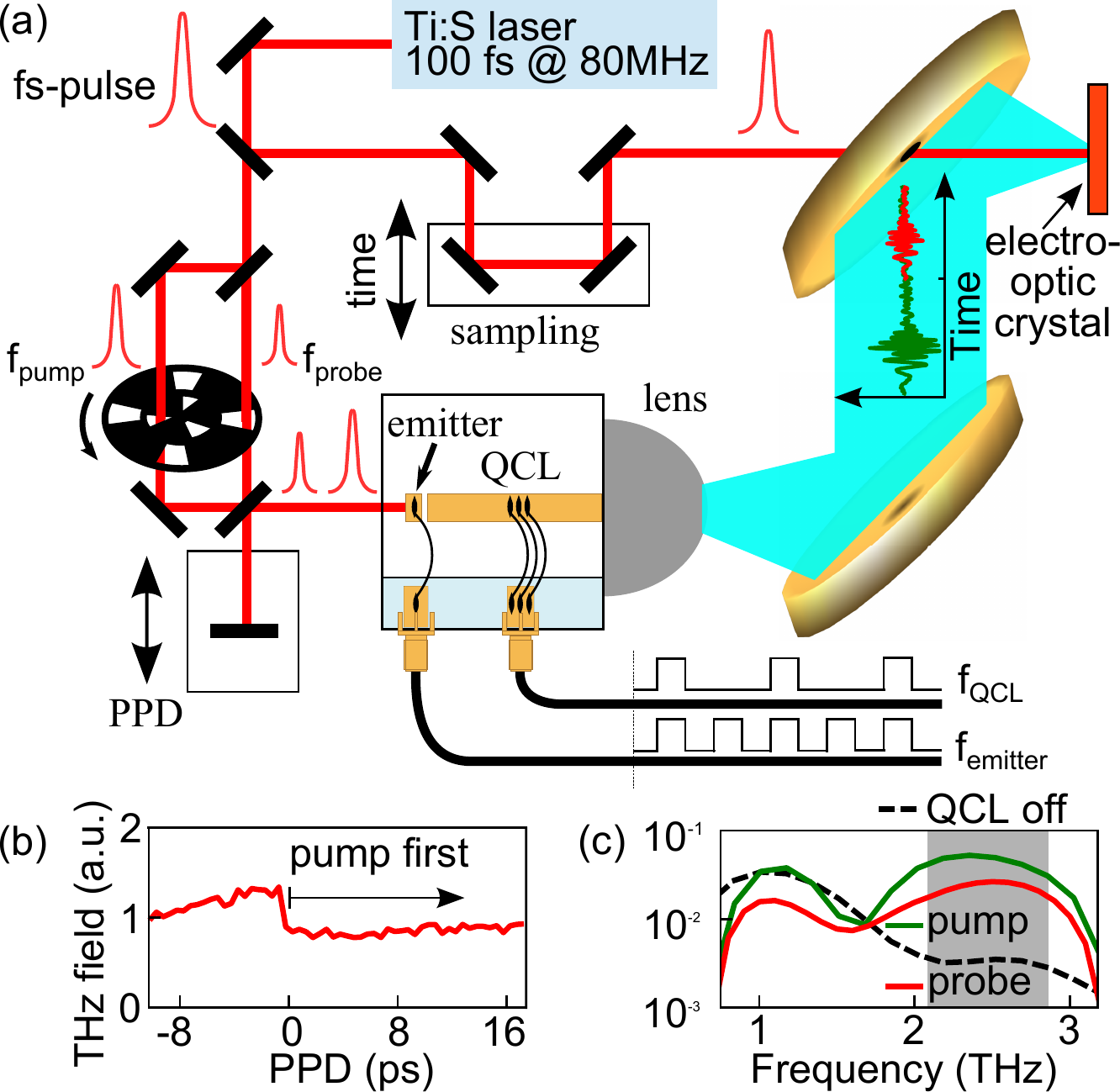}
\caption{THz-pump/THz-probe spectroscopy for the assessment of the gain recovery time in a THz QCL: (a) schematic of the optical setup and the modulation scheme; (b) amplitude of the THz-probe, which is generated in the integrated emitter for the THz-pump at 0$\,$ps, is plotted as a function of the pump-probe delay time (PPD); (a) and (c) THz pulses and the spectral content of the THz pulses after the propagation through the QCL section.}
\label{fig:fig1}
\end{figure}

In this Letter, we present the measurement results on the gain dynamics of a broadband heterogeneous THz QCL active region.\cite{Turcinkova2011} Using a THz-pump/THz-probe technique, the time evolution of the gain depletion is evaluated. We show that the gain recovery is in the order of 34$\,$-$\,$50$\,$ps when the QCL active region is in steady-state and driven above lasing threshold. The gain recovery time has been estimated for the whole dynamic range of the laser and for the complete temperature window of the laser operation. The observed gain behavior is modeled and satisfactorily explained using a rate equation model of the quantum cascade heterostructure.

The active region of the studied THz QCL consists of an interlaid stack of three different quantum cascade heterostructures (centered @ 2.3$\,$THz, 2.6$\,$THz, 2.9$\,$THz, respectively; for further details see Ref.\cite{Turcinkova2011}). This THz heterogeneous active region provides a gain broader than 1$\,$THz \cite{doi:10.1063/1.4901316} and its spectral flexibility was demonstrated by THz laser devices with an octave spanning emission.\cite{Roesch2014} The THz QCL under test features a 75$\,$\textmu m wide double-metal waveguide with a length of 2930$\,$\textmu m.
We have employed THz time domain spectroscopy, which provides direct access to the optical gain of the QCL via the amplification of the injected broadband THz pulses.\cite{Kroell2007} The photo-conductive THz emitter is integrated on the QCL chip using a coupled cavity device geometry.\cite{doi.org/10.1364/OE.19.000733} To enable the THz-pump/THz-probe technique, the conventional THz time domain setup was extended to two independent near-infrared (NIR) pump beams, one with a fixed and the other with a variable optical path length that we define as ``THz-pump'' and ``THz-probe'' path, respectively. The delay time between the preceding THz pump pulse and the subsequent THz probe pulse is called pump-probe delay (PPD). A schematic of the optical setup is depicted in Fig.~1(a) and indicates also the used modulation scheme. The THz emitter and the THz QCL are driven with electrical pulses at the modulation frequencies $f_{\text{emitter}}\,$=$\,$20$\,$kHz and $f_{\text{QCL}}\,$=$\,$10$\,$kHz, respectively, similarly to the standard TDS measurement procedure of QCLs.\cite{Kroell2007} In addition, the THz-pump and THz-probe pulse trains are modulated mechanically at the frequencies $f_{\text{pump}}\,$=$\,$450$\,$Hz and $f_{\text{probe}}\,$=$\,$630$\,$Hz, respectively. This modulation provides us simultaneous access to the fraction of the transmitted signal that is caused by the THz-pump pulse, the THz-probe pulse and the presence of both pulses. 
All four necessary modulation frequencies are generated by RF generators locked to a common frequency reference. 
The signal of the electro-optic detector is demodulated using synchronized lock-in amplifiers at the frequencies $ \left( f_{\text{emitter}} - f_{\text{QCL}} + f_{\text{pump}} \right)$, $ \left( f_{\text{emitter}} - f_{\text{QCL}} + f_{\text{probe}} \right)$ and $ \left( f_{\text{emitter}} - f_{\text{QCL}} + f_{\text{pump}} + f_{\text{probe}} \right)$, which allows us to measure simultaneously the pump, the probe and the modulation of the probe due to the presence of the preceding pump (for $\text{PPD} \geq 8\,$ps), respectively.

Since the THz emitter is integrated into the QCL waveguide,\cite{doi:10.1063/1.3280038} we have checked the linearity of the THz generation process with respect to the NIR pump power. We used NIR average powers up to 68$\,$mW and 55$\,$mW to generate the THz-pump and THz-probe pulses, respectively. The panel of Fig.~1(b) shows the intensity of the generated THz-probe pulses directly scattered from the emitter (so called air-side pulses\cite{doi:10.1063/1.3280038}) for different time delays with respect to the THz-pump pulse. The amplitude of the THz-probe pulses shows $\sim\,$10$\,$\% reduction when the probe pulse is generated after the THz-pump pulse, and this reduction persists almost constant in the whole time window of interest -- 8$\,$ps to 60$\,$ps. This behavior of the probe indicates a long-lasting partial saturation of the THz emitter by the NIR pump pulse. This partial saturation affects our measurement of the time evolution of the QCL gain depletion just as a constant offset of the modulation signal amplitude and is accounted during the data evaluation.
\begin{figure}[h!]
\centering\includegraphics[width=0.85\linewidth]{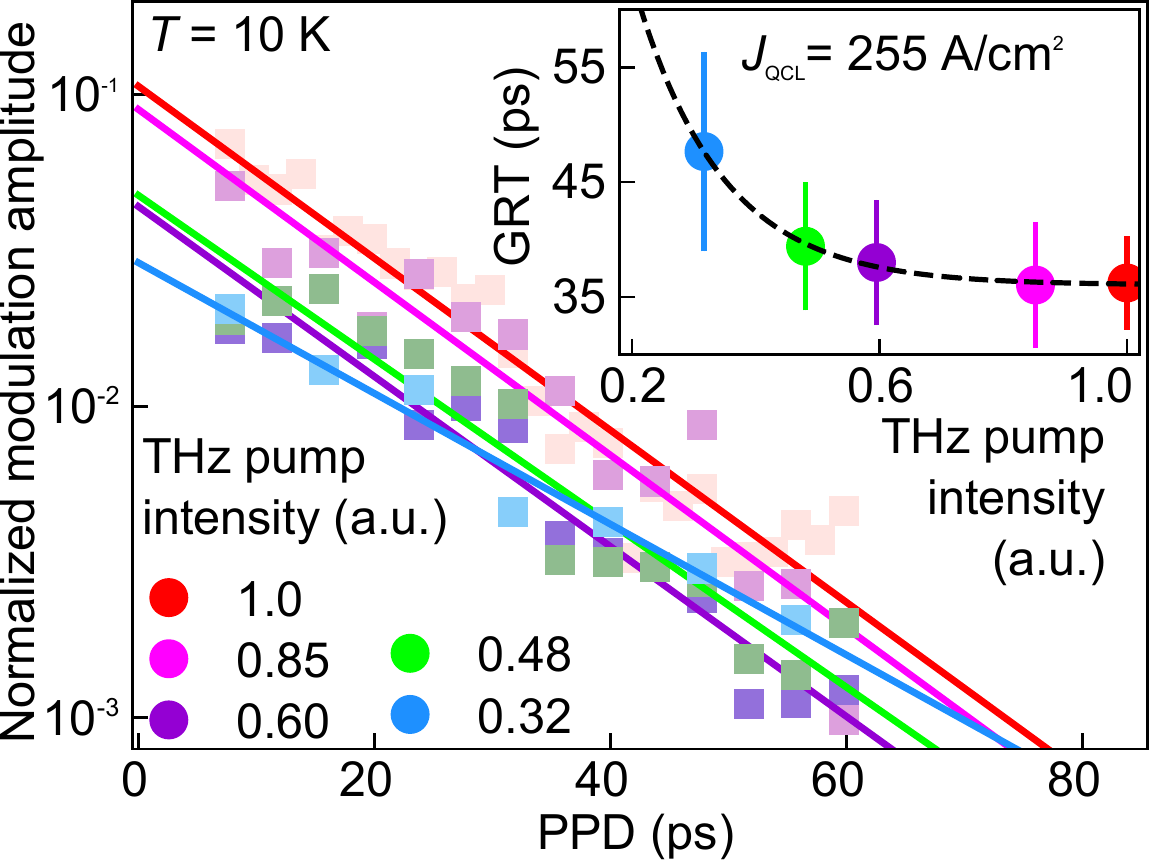}
\caption{The gain recovery in a THz QCL with a heterogeneous active region at several different THz-pump pulse intensities related to the maximum -- for NIR pulses with 62$\,$mW power focused on the emitter. (dots -- GRT; vertical bars -- error margins; dashed line -- a guide for the eye)}
\label{fig:fig2}
\end{figure}
Typical shapes of THz pulses transmitted through the QCL section and their spectra are shown in Fig.~1(a) and 1(c), respectively. 
The QCL is usually transparent for frequencies below $\sim\,$1.3$\,$THz and thus serves as calibration for the THz pump pulse. The high frequency part of the spectrum ($> \,$2$\,$THz) is shaped by the gain curve of the QCL and its amplitude scales with the bias current density of the laser.\cite{doi:10.1063/1.4901316} The intermediate frequencies around 1.5$\,$THz are partially suppressed due to an absorption at the low frequency side of the QCL gain curve -- a typical feature of QCHs observed previously.\cite{MMartlThesis2011, doi:10.1063/1.3670050} The spectra of the THz-pump and THz-probe pulses are similar and their amplitudes scale linearly with the NIR pulse amplitude indicating that the THz emitter is operated in the linear regime and the perturbation due to injected THz pulses is not frequency selective.
In order to achieve a good signal-to-noise ratio, we used the integral spectral amplitude between $2.2\,$THz and $2.8\,$THz, see the shadowed area in Fig.~1(c), to extract the recovery dynamics of the gain.

Figure 2 shows the amplitude of the modulation of the THz-probe due to the preceding THz-pump normalized to the probe amplitude in the considered pump-probe time delay window between 8 and 60$\,$ps for five different intensities of the THz-pump.
The THz-probe modulation is decaying exponentially with the time delay due to the gradual recovery of the gain depleted by the THz-pump pulse as indicated by the straight lines fitting the measured data plotted in log scale. The slope of these lines gives the GRT of 36$\,\pm\,$2$\,$ps for the QCL driven close to the maximum output power
and for a wide range of THz-pump amplitudes. The gain recovery dynamics gets remarkably slower with decreasing THz-pump amplitude, which indicates that we are departing from the operation regime dominated by the density of photons in the laser cavity, in which the stimulated emission accelerates the gain recovery. 
A similar pump-intensity induced dependency of the GRT was also observed in Ref.~\cite{BaconThesis2017}. 
In further investigations we have limited ourselves to a regime with a large photon density perturbation as such a regime resembles the conditions for laser pulsing.

\begin{figure}[!ht]
\centering\includegraphics[width=1.00\linewidth]{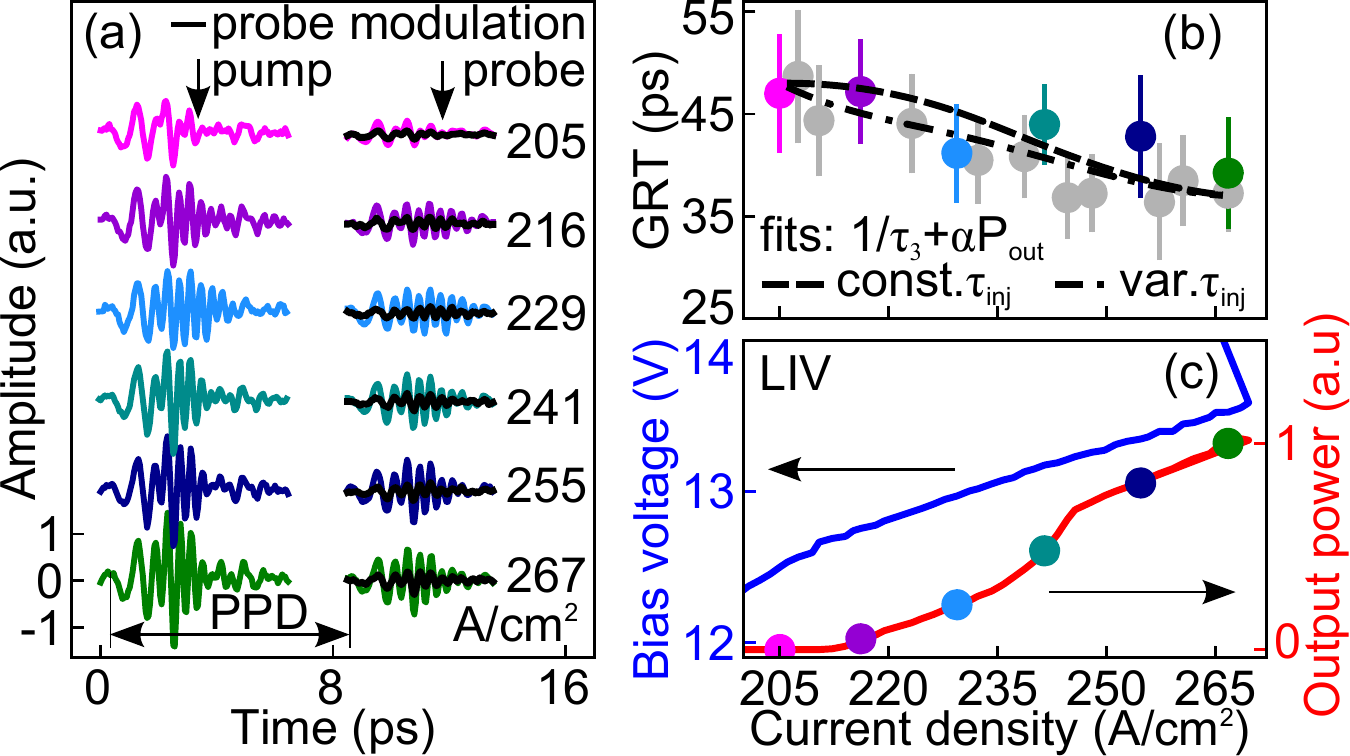}
\caption{The gain recovery for different driving conditions of the QCL: (a) selected time-domain THz-pump/THz-probe data for a pump-probe delay (PPD) of 8$\,$ps (pulses are vertically offset for clarity); (b) gain recovery time as a function of the driving current density (gray dots -- the time-domain data is not shown in (a)); (c) THz output power and QCL bias voltage as a function of driving current density.}
\label{fig:fig3}
\end{figure}

We have investigated the gain recovery time in the whole driving current range above the lasing threshold for a current density of 200$\,$-$\,$265$\,$A/cm$^2$ (Fig.~3). 
We observe that the gain recovery time -- extracted from the time traces in Fig.~3(a) --  changes monotonically between 48$\,$ps and 36$\,$ps. 
When we correlate the measurements of the driving current dependent gain recovery with the driving current dependent output power of the QCL (Fig.~3(b) and 3(c)), we find that the higher photon density (i.e. higher THz output power) in the QCL is associated with a shorter gain recovery time. 
Furthermore, we have also studied the temperature dependence of the gain dynamics (see Fig.~5).
When the QCL is driven close to the point of maximum output power, the GRT values exhibit two maxima as a function of the operating temperature -- one in the range of 35$\,$-$\,$50$\,$K and another one for temperatures above 60$\,$K.
To find a qualitative explanation of the observed gain recovery dynamics in our driving current-dependent and temperature-dependent studies, we model the electron transport in the QCH (Fig.~4) using the standard three-level rate equation model for QCHs.\cite{FaistBookQCL2013} As we focus ourselves on the laser operation conditions above the lasing threshold, we neglect the spontaneous emission rate in our model. The GRT is determined from the temporal evolution of the upper laser level  population and given by
\begin{equation}
 \frac{\partial n_{\text{s},3}}{\partial t} = \frac{J}{q_{\text{e}}} - g_{\text{c}} \left( n_{\text{s},3} -n_{\text{s},2} \right) n_{\text{ph}} - \frac{n_{\text{s},3}}{\tau_3} 
\label{eq:one},
\end{equation}
where $n_{\text{s}3}$, $n_{\text{s}2}$, and J are the sheet densities of the upper $\left( \Ket{3} \right)$ and the lower $\left( \Ket{2} \right)$ lasing levels, and the bias current density; $n_{\text{ph}}$, $g_{\text{c}}$, and $\tau_3$ are the density of photons, the coefficient for stimulated emission, and the electron lifetime of the upper lasing level for the empty cavity, respectively. In general, all coefficients are dependent on the applied electric field, which modifies the overlap factors of the wavefunctions, and on temperature if phonon scattering assists the transition between the two particular energy levels. The solution of the differential equation (1) for a small perturbation of the upper lasing state population by a THz-pump pulse is given by
\begin{equation}
 \Delta n_{\text{s},3}(t) \propto \exp\left[  -t \cdot \left(    \tau_3^{-1} + g_{\text{c}} n_{\text{ph}}  \right)  \right] 
\label{eq:two}.
\end{equation}
This expression states that the population of the upper lasing state recovers faster when the photon density increases or the electron lifetime of the upper lasing level decreases. This simple equation will allow us to explain our measurement results qualitatively and quantitatively.

First, for our driving current dependent study above lasing threshold we assume that there is only a weak dependence of the upper lasing level lifetime   (non-radiative) on the QCL bias and the main contribution is due to stimulated emission. This allows us to fit the GRT data in Fig.~3(b) (as indicated by a dashed line) with Eq.(2) using the QCL output power data in Fig.~3(c). An even better fit can be obtained when we assume that the non-radiative lifetime is changing at the onset of lasing (dash-doted line). In the case of a bias dependent injection, the injection rate has to be introduced into the non-radiative lifetime, according to $J/q_e \approx (n_{\text{s,1}} - n_{\text{s,3}})/\tau_{\text{inj}}$. 
We find a factor of $\sim\,$3 between the non-radiative $ \left(  1/\tau_3   \right)  $ and radiative $ \left(  g_{\text{c}} \cdot n_{\text{ph}}   \right)  $ decay rates (i.e. stimulated emission rate) from the upper lasing level. This information provides valuable insight into the internal processes of the studied THz QCL and indicates that only about$\,$1/4 of the electrons injected into the upper lasing state contribute to the THz generation. Considering the estimated efficiency of the lasing transition together with the ratio of the THz-photon energy to the voltage drop per quantum cascade (52$\,$meV) we find a theoretical upper bound for the wall-plug efficiency of about 5$\,$\% for the studied THz QCL.

\begin{figure}[!ht]
\centering\includegraphics[width=0.990\linewidth]{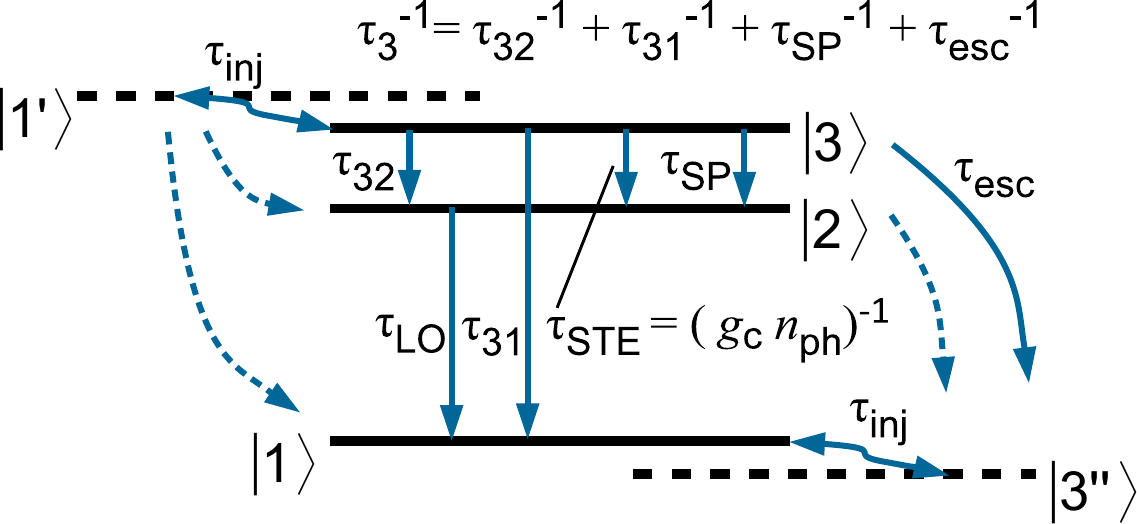}
\caption{Rate equation model for the THz quantum cascade heterostructure design (dashed-line arrows -- direct scattering from/into injector states that are neglected in the considered model).}
\label{fig:fig4}
\end{figure}

\begin{figure}[!ht]
\centering\includegraphics[width=\linewidth]{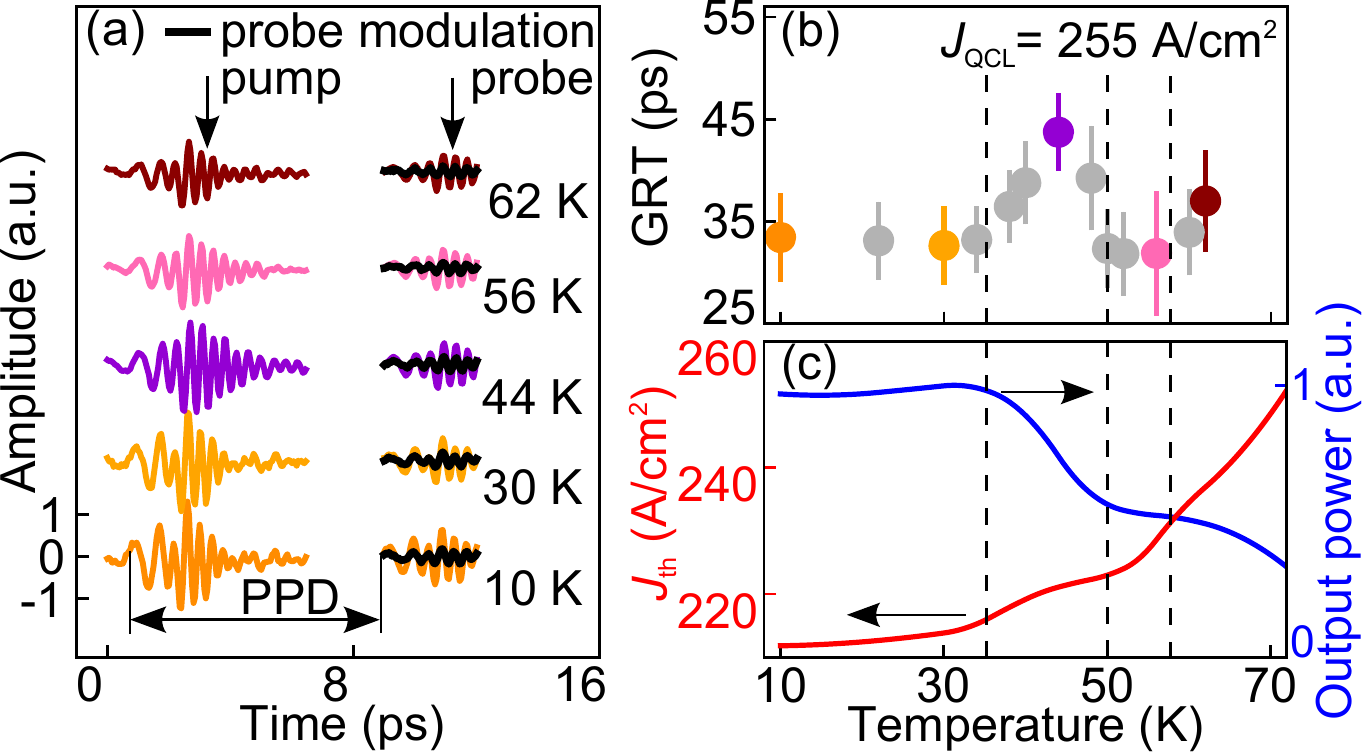}
\caption{Temperature dependence of the gain recovery of the THz QCL: (a) selected time-domain THz-pump/THz-probe data (pulses are vertically offset); (b) Observed gain recovery time (gray dots -- data not displayed in (a)); (c) The threshold current $J_{\text{th}}$ and THz output power as a function of temperature.}
\label{fig:fig5}
\end{figure}

Second, we focus on the changes of the gain recovery dynamics at different operation temperatures of the QCL. The measurement is performed in the temperature range of 10$\,$K to 72$\,$K, for which the THz gain in the studied laser has been previously reported to be constant.\cite{doi:10.1063/1.4901316} Figure~5(a) and 5(b) show THz-pump and THz-probe pulses at several selected laser operation temperatures, and the gain recovery times observed at these temperatures, respectively. The gain recovery dynamics is constant up to about 35$\,$K, where it starts to slow down, and the fast dynamics recovers back at about 50$\,$K. The increased gain recovery time in the given temperature window (35$\,$K to 50$\,$K) correlates with the onset of a gradual decrease of the QCL output power (Fig.~5(c)). This is in agreement with our model, which states that a decrease of the photon density can slow down the gain recovery. There are four major phenomena responsible for the reduction of the QCL power with temperature, namely the increase of waveguide loss, which is a minor effect in the temperature range of 10$\,$-$\,$50$\,$K, the thermally activated back-filling of the lower lasing state $\Ket{2}$, and the scattering of electrons from the injector directly to the level $\Ket{2}$. The fourth one is the escape of electrons from quantum wells to the continuum or to the extraction level $\Ket{1}$ that leads to a faster dynamics on the level $\Ket{3}$ and can be excluded for now. Therefore, a tentative explanation for the observed temperature dependence of the GRT relies on the THz gain reduction due to back-filling of level $\Ket{2}$ from the extraction level $\Ket{1}$ and/or the scattering of electrons from the injector directly to the level $\Ket{2}$ leading to a reduced population difference $\left(n_{\text{s,3}} - n_{\text{s,2}} \right)$. The original fast recovery dynamics recovers back as the operation temperature approaches 50$\,$K and is accompanied by a break in the QCL output power degradation. Such behavior indicates a compensation of the level dynamics that are lost due to the lower photon density.
The second slow-down of the gain recovery correlates with the QCL output power roll-off above 60$\,$K and with the rising lasing threshold current density. It is attributed to the dominant electron scattering from the injector to levels $\Ket{2}$ and $\Ket{1}$. Therefore, in this area the population dynamics of the upper lasing level $\Ket{3}$ is controlled by the slow electron supply from the injector and in the used model the current density $J$ has to be replaced by the carrier tunneling rate 
$\Delta n / \tau_{\text{inj}}$, analogous to the driving current dependent study, where we compare the model with the unique effect of the intracavity photon density with the extension of a variable upper state lifetime (including a variable injection rate, see Fig.~3(b)).
This explanation goes beyond the simple model considered by Bachmann \textit{et al.} for the gain roll off in which the simple degradation of the population inversion $\left(n_{\text{s,3}} - n_{\text{s,2}} \right)$ is considered.\cite{doi:10.1063/1.4901316} Our experimental data point directly towards a reduced injection efficiency of electrons from the injector to the upper lasing state $\Ket{3}$. This phenomenon is expected, since at 60$\,$K the dynamic range of the laser is highly reduced and the operating point on the I-V curve is very close to the rollover point where the electrical stability and also the injection is not optimal anymore.

The gain recovery time between 34$\,$ps and 50$\,$ps that we determined by studying BTC QCHs with resonant-phonon extraction above threshold is generally longer than the GRT of 15$\,$-$\,$26$\,$ps found in other similar studies of THz BTC QCHs.\cite{doi:10.1063/1.4942452, Freeman2013, Markmann:17,  BaconThesis2017} 
The theoretical calculation of a bound-to-continuum quantum cascade at 3.74$\,$THz also indicated a quite short upper state lifetime of $\sim\,$12$\,$ps.\cite{doi:10.1063/1.2759271} 
However, all those results have been obtained for the classical BTC design of THz QCH. In those designs the lower lasing level, which is the top-most energy level of the injector miniband constituted by five or more states, is coupled via electron thermalization to the upper lasing level of the next cascade, which is at the bottom of the injector. A substantially longer GRT of 50$\,$ps has been reported by Green \textit{et al.} for a similar BTC design at 3.1 THz measured using THz pump pulses from a free electron laser.\cite{Green2009} Their result can be understood as a very slow recovery of the strongly depleted gain (i.e. burned spectral holes) typical for narrow-band pump pulses. This is obviously not the scenario in our case, so the rather long gain recovery time presented in this letter is an unique feature of the studied THz QCH design.
Finally, by comparing the effect of the photon density and operation temperature on the gain recovery dynamics, we find that similar tendencies were observed for THz QCL with BTC design \cite{BaconThesis2017} and for the BTC with resonant phonon extraction design studied in this letter.
Bacon\cite{BaconThesis2017} also observed a temperature induced slow down of the gain recovery dynamics, and is assigning it to the temperature dependence of the carrier-carrier and carrier-phonon scattering that are influencing the coherence time of the resonant tunneling injection.

In conclusion, we have performed THz-pump/THz-probe measurements to determine the gain recovery dynamics in a broadband THz QCL with an active region based on a bound-to-continuum optical transition with an optical phonon assisted extraction. From the temperature and intracavity power dependence measurement we determined a gain recovery time in the range of 34$\,$-$\,$50$\,$ps. The short time limit is observed at low operation temperatures and high THz output powers of the laser, while the rather slower recovery dynamics is observed when the phonon scattering of electrons is compromising the electron transport via the quantum cascades (i.e. at higher operation temperature or low driving current). The measurement results were analyzed using the standard three-level rate equation model of QCHs, which provides a relation between intracavity photon density and the gain recovery. 
Furthermore it allows to explain the role of electron-phonon scattering in the efficiency to pump the upper laser level and to extract the electrons from the lower laser level with increasing temperature.

The authors CGD, DB, KU and JD acknowledge financial support from the Austrian FWF in the framework of the Doctoral Schools ``CoQuS'' (W1210) and ``Solid4Fun'' (W1243), the SFB projects ``NextLite'' (F4902) and ``DIPQCL'' (P30709) and the Austrian GMe. The authors GS, MB and JF acknowledge the SNF support under Contract No. 200020-165639 and H2020 European Research Council Consolidator Grant CHIC 724344.

Author contributions: JD and KU devised the experiment, JD, DB, and CGD built the setup; DB fabricated THz QCL lasers, CGD performed the measurement and evaluated the data; JD and CGD did the modeling; GS, JF and MB designed and grew the THz QCL; all authors have discussed the results and contributed to the text of the manuscript.

\providecommand{\noopsort}[1]{}\providecommand{\singleletter}[1]{#1}%
%


\end{document}